\documentclass{article}

\usepackage{arxiv}
\usepackage{array}
\usepackage[utf8]{inputenc} 
\usepackage[T1]{fontenc}    
\usepackage{hyperref}       
\usepackage{url}            
\usepackage{booktabs}       
\usepackage{amsfonts}       
\usepackage{nicefrac}       
\usepackage{microtype}      
\usepackage{lipsum}
\usepackage{graphicx}
\graphicspath{ {./images/} }
\usepackage{subcaption}

\newcommand{\etal}{\textit{et~al.}}

\title{Who’s Watching You Zoom?\\  Investigating Privacy of Third-Party Zoom Apps}

\author{
 Saharsh Goenka \\  
  Arizona State University\\
  Tempe, AZ \\
  \texttt{sgoenka1@asu.edu} \\
   \And
 Adit Prabhu \\
  Arizona State University\\
  Tempe, AZ \\
  \texttt{adprabh2@asu.edu} \\
   \And
 Payge Sakurai \\
  Arizona State University\\
  Tempe, AZ \\
  \texttt{psakurai@asu.edu} \\
   \And
 Mrinaal Ramachandran  \\
  Arizona State University\\
  Tempe, AZ \\
  \texttt{maramach@asu.edu} \\
   \And
 Rakibul Hasan \\
  Arizona State University\\
  Tempe, AZ \\
  \texttt{rakibul.hasan@asu.edu} \\
}

\begin{document}
\maketitle
\begin{abstract}
Zoom serves millions of users daily and allows third-party developers to integrate their apps with the Zoom client and reach those users. So far, these apps' privacy and security aspects, which can access rich audio-visual data (among others) from Zoom, have not been scientifically investigated. This paper examines the evolution of the Zoom Marketplace over one year, identifying trends in apps, their data collection behaviors, and the transparency of privacy policies. Our findings include worrisome details about the increasing over-collection of user data, non-transparency about purposes and sharing behaviors, and possible non-compliance with relevant laws. We believe these findings will inform future privacy and security research on this platform and help improve Zoom's app review process and platform policy.
\end{abstract}


\section{Introduction}

Zoom experienced a dramatic rise in popularity during the Covid-19 pandemic, and has become the dominant tool for remote conferencing with more than 300 million daily active users~\cite{zoom2021annual}. In the US, Zoom has also become the official tool in many public and private organizations, including education and healthcare institutes, as well as business entities. Recently, Zoom has turned into a platform, rather than just a tool, that hosts many other applications (apps) offering additional services. These tools, which are available in the Zoom marketplace, are most commonly provided by third-party developers.

Like apps in other platforms (e.g., Android), Zoom apps access user data, inevitably raising privacy and security concerns. These issues have been extensively studied in many other domains like mobile platforms and voice assistants (see~\S~\ref{sec:app-trend-review}). However, by nature of the services provided, Zoom presumably creates a lot more audio-visual data, which are rich sources of a lot of other information about the users, all of these can potentially be accessible to third-party developers. Zoom's rise was also followed by the so-called AI-boom, where access to predictive or generative models became commonplace. This trend is also observed in the Zoom marketplace, with a lot of apps utilizing ``AI'' (that not only worsens existing privacy risks but also creates new ones~\cite{leeDeepfakesPhrenologySurveillance-2024}) to provide features. Zoom has faced backlash over AI-based features such as emotion recognition~\cite{zoom-emotion-backlash} and its policy to use customer data to train AI-models~\cite{zoom-ai-training-backlash}. Although Zoom has backed off those plans, the marketplace hosts many third-party apps that provide similar features. 

As Zoom is becoming increasingly popular in remote education and healthcare services, marketplace apps presumably serve students (including minors) and healthcare consumers. Personal data created in these contexts are subject to additional regulations (see~\S~\ref{sec:background}); while Zoom itself is compliant with relevant laws, compliance is not automatically extended to third-party apps.

All of the above issues create an urgency and strong motivation to study the app marketplace and investigate privacy and security aspects of the available apps. 
In this paper, we present a year-long study assessing the evolution of the Zoom app marketplace. 
Our study makes several important contributions to understanding the Zoom app marketplace. We conducted a longitudinal analysis that tracked changes in the Zoom marketplace over one year, from December 2023 to December 2024, and created a comprehensive dataset  
containing 97,194 snapshots of Zoom Marketplace apps. 

The analysis of this dataset reveals growth trends in third-party applications and shifts in popular categories. We then examine the permission requests of these applications and associated privacy risks, documenting a notable rise in excessive data permission requests, particularly among newer applications. We observe potential misuse of these categories, with instances where apps encompass multiple categories, possibly to broaden their reach, often without providing relevant functionalities.

Furthermore, we analyzed privacy policies at different time points to understand trends in disclosing data collection and usage. We uncover issues with transparency, such as vague data collection statements, and omission of purposes for data collection. We also found that only a small number of privacy policies indicate that they comply with relevant laws (e.g., FERPA~\cite{ferpa}). We discuss the privacy, safety, and ethical implications of these findings. 


\section{Background and literature review}

\subsection{Zoom apps and marketplace}\label{sec:background}
Zoom launched the Marketplace on October of 2018 for third-party developers to publish apps that will operate within Zoom client for desktop/laptop and mobile platforms. Third-party apps can interact with Zoom and access data in several ways. For contextual data, i.e., data relating to the current environment, can be accessed via the Zoom Apps JS SDK~\cite{zoom-js-sdk}. The Zoom Apps SDK is a JavaScript library that facilitates communication between the marketplace application and the Zoom client. Server-side data and events from the Zoom account, including calendar information, meeting reports, cloud recordings, and account data, are accessed via Zoom REST APIs \cite{zoom-api}. Media Streams and meeting chat data are accessed via Zoom Meeting SDK \cite{meeting-sdk}, which uses Meeting Bots to connect to meetings as a participant and generate or process the media data streams. Zoom encourages developers to minimize data collection and use, and the use of granular scopes to be specific in terms of what data they need~\cite{zoom-scopes}.

\paragraph{Regulatory compliance.}
Apps in certain categories may be subject to additional laws. In their app review guideline~\cite{zoom-app-review-guide}, Zoom noted the increasing popularity of third-party apps among students (including minors) and guided how developers should guard data to comply with FERPA~\cite{ferpa} and COPPA~\cite{coppa}, which are the US federal regulations that dictate the collection and use of educational records and data about children. Additionally, personal health information (PHI) is protected under HIPAA~\cite{hipaa-definition}. Unfortunately, the US does not have a comprehensive federal privacy law, and the sector-specific ones mentioned above may not apply to all entities (such as some private companies). Developers can comply with those regulations either directly, or by entering into a business contract with other covered entities. For example, an education (or healthcare) app may provide service to public K-12 schools (or hospitals), which are covered entities, and can create contracts (commonly known as a BAA or  Business Associate Agreement) to comply with regulations. Zoom allows third-parties to enter into BAAs with Zoom, which directly complies with all of the above laws. However, when such apps are used by individuals (e.g., teachers using apps in classrooms~\cite{kelso2025investigating}), compliance cannot be enforced, raising concerns about severe privacy breaches.


\subsection{App marketplace trend analysis}\label{sec:app-trend-review}
Researchers have studied privacy and security practices on app marketplaces for many platforms. For example, Wang~\etal~\cite{wangUnderstandingEvolutionMobile_2019} collected three snapshots of the Google Play store spread across more than three years to study how the app ecosystem evolved. Their longitudinal analysis tracked over 160,000 apps and revealed concerning trends. Many apps were requesting additional permissions without adding corresponding functionality, permissions requests increased alongside app popularity, and privacy policy accessibility decreased over time. Besides general trends in the marketplace, their study identified issues with data use and privacy policy declaration and the existence of malicious apps and developers. Edu~\etal measured the trend of the Alexa marketplace for three years to investigate the evolution of this ecosystem, provided clarity on data disclosure practices, and identified skills with issues impacting their security and privacy~\cite{eduMeasuringAlexaSkills2022}. Their study of over 90,000 Alexa skills revealed that only 24.2\% of skills have privacy policies, and skills increasingly requested more permissions over time, with the most significant requests about location and profile information.

Zhang ~\etal~\cite{zhang2024lookgptappslandscape} studied the evolution of GPT Marketplace for 10 months and found that, unlike mobile apps, GPT applications rapidly evolved their capabilities and permission requests in this early marketplace stage. 
They also identified significant vulnerabilities in GPT app configurations, with system prompts, knowledge file names, and file contents successfully extracted from 90\%, 88\%, and 12.7\% of apps posing risks to creators' data. Similarly, Zhang ~\etal~\cite{wechat} studied the WeChat Mini-Programs marketplace, which hosts over one million applications within China's leading social platform. 
Over 50\% of mini-apps in the education, business, lifestyle, utilities, and gaming categories were removed from the marketplace frequently, suggesting potential quality or policy violation issues. 

While Android and WeChat marketplaces have been significantly researched, Fuqi ~\etal~\cite{ios-removed} conducted the first comprehensive study of removed apps in the iOS App Store through 1.5 years of daily marketplace snapshots. Their analysis of over 1 million removed apps revealed that app removal follows cyclical patterns, with large-scale removals happening monthly. About 5\% of removed apps were previously popular, ranked in top-1500, and over 73\% of developers who had apps removed also had all of their released apps eventually removed from the store as well. 

These marketplace studies demonstrate a consistent pattern across platforms: increasing permission requests over time, often without corresponding functional improvements, and varying degrees of policy enforcement.

\subsection{App permission evolution analysis}
The first comprehensive study on permission evolution in the Android platform was conducted by Wei ~\etal~\cite{weiPermissionEvolutionAndroid-2012}. They examined multiple Android releases over three years, analyzing 237 third-party apps(1703 versions spanning 3 years) and 346 pre-installed apps(1714 versions). Their research revealed several concerning trends. The number of permissions defined in Android apps grew over time, with Dangerous-level permissions(permissions regarding personal info, accounts, etc) being the most frequent category; permission additions also made up the majority of app evolution, with 90.46\% of permissions changes being additions rather than removals; an increasing percentage of apps (44.8\%) violated the principle of least privilege by requesting permissions they apparently did not use; and pre-installed apps had access to higher-privileged permissions, creating significant security and privacy risks for users who had no say on installing them. 

Calciati and Gorla later conducted similar analyses on a larger set of Android apps and reported the same trends, e.g., apps ask for more permissions over time~\cite{calciatiHowAppsEvolve-2017}. Their study of 14,000 apps showed that 49\% of apps increased their permissions request in subsequent releases. Taylor and Martinovic studied the evolution of the so called dangerous permissions on the Android platform and reported that apps increase those permissions in subsequent release often without adding new functionality~\cite{taylor2016longitudinal}. 
While a majority of the studies on app permissions focused primarily on Android permission systems, Garg and Baliyan \cite{AppPermissions} conducted a comprehensive comparative analysis between Android and iOS security models. In their study, they went in-depth on their respective permission systems and revealed fundamental differences in how these platforms handle permissions. 
Their analysis of vulnerability data demonstrated that Android had a higher percentage of vulnerabilities related to permission abuse with 96\% of privilege escalation vulnerabilities compared to 4\% in iOS. This aligns with Taylor and Martinovic's~\cite{taylor2016longitudinal} findings that dangerous permission requests increase over time, suggesting that Android's permission model may contribute to permission creep. 
The collective findings from these permission evolution studies highlight concerning trends across mobile platforms, as well as the role system architecture plays in these evolutions. Regardless of platform, apps tend to accumulate permissions over time and with increases in popularity. This pattern is particularly problematic in platforms like Android, where architectural decisions enable developers to easily modify permissions, and in environments like Zoom marketplace where third-party apps may access sensitive data in regulated fields such as healthcare and education. Our analysis of Zoom marketplace builds upon these methodological approaches to examine whether similar patterns of permission evolution occur in this relatively unexplored ecosystem, and what implications they might have for user privacy. 
\subsection{Privacy policy analysis}
App developer's privacy policy is the primary source documenting data collection and sharing practices and has been the subject of research from multiple perspectives, including automated analysis and summarization of policy documents~\cite{cui2023poligraph}, checking the consistency between stated policy and actual behaviors~\cite{privacy-policy-vs-reality} ~\cite{android-privacy-complicianc-extra?}, evolution of policy documents~\cite{privacy-policy-over-time}, as well as checking compliance with regulatory measures~\cite{xiangPolicyCheckerAnalyzingGDPR_2023}. Analyzing trends in website privacy policies over more than two decades,  Amos~\etal~\cite{amosPrivacyPoliciesTime-2021a} reported concerning findings, such as failure in disclosing data collection and tracking practices and continuous expansion of policy documents while their readability worsening over time, making it difficult to get meaningful information and raising concerns about informed consent from users. 
Alamri, ~\etal~\cite{ios-policy-adherence}, looked at the meta data for 2 million apps to see how many apps provide links to valid privacy policies and if those links actually work. The results showed that only 58.5\% of apps had a privacy policy link, and only 38.4\% of that 58.5\% had actual valid privacy policies. Another study, Zimmeck ~\etal~\cite{zimmeck2017automated} finds that in a study of 17,991 free apps on the Android store, 71\% of apps that lack a privacy policy should have one. 
\section{Methods}
\subsection{App marketplace data collection}\label{sec:data-collection}

From December 2023 to December 2024, we collected data about apps on the Zoom marketplace. First, we crawled the marketplace directory to compile an exhaustive list of URLs to individual app pages and then crawled those pages to collect app details and privacy policies. The crawling was executed at the beginning of each week (Sundays at midnight) to maintain data currency while minimizing disruption to the marketplace's operations. Each complete crawling session of the Zoom Directory, the individual pages of the app, and the associated privacy policies typically took approximately four hours on average, with a 10-second delay between consecutive requests to avoid exhausting the server.

For this purpose, we developed a specialized crawler based on the Puppeteer library~\cite{puppeteer} and a parser, and periodically updated them to handle technical issues and changes in the marketplace and app details page format. The first change was the introduction of app categories in the marketplace in March 2024, which subsequently underwent additional changes, such as different locations of category data on the page and how they were presented. Additionally, the privacy policy link on app pages was moved from the bottom to the top of the page. We addressed these issues by updating the parser to check the new location and using keywords to search for the link rather than solely relying on CSS tags to locate the element. We likewise updated the parser to address changes in how and where app scopes were listed. Some of the technical issues we faced were the unavailability of the server at times, slow loading of pages leading to timeout errors, and invalid or non-existent links to other pages and documents (particularly privacy policies).

We created another crawler and parser to handle a major change in the app category listing after May 2024. Previously, all categories under which an app was listed were included on the app details page. However, after May 2024, only the first category was listed, and additional categories were loaded and made visible after hovering over the category-listing area. Triggering this hovering action automatically could not be done reliably. Thus, we created a crawler that periodically visited all web pages that listed app names under specific categories (there were 32 categories in total). We also made a parser to extract app names and other details for post-processing.

Despite these technical challenges, we ensured reliability in the data collection process through extensive logging, robust error-handling mechanisms, and recovery steps for any lost data. The crawlers and parsers logged every request, as well as errors and exceptions they faced. The project lead also would receive email notification if they had to halt operation, and manually reviewed logged messages and updated the data collection framework as needed. To prevent data loss, the crawler saved all HTML pages so that even if the parser fails (e.g., due to a new change in page format), we could adapt the parser and recover data from the saved pages. 



We also implemented automated verification steps to enhance data accuracy and completeness. For example, after each cycle, the system compared the total number of apps listed in the marketplace directory with the number of apps for which data had been collected, ensuring that data was gathered for all available apps. There was also an edge case where an app could be created or deleted during the 4-hour data collection window. In such instances, the Data framework documented these changes in the email log sent to researchers, allowing for manual verification and recovery. The parser checked for any null values in essential data fields for each app, such as the app developer and the privacy policy. If a null value was detected, the app would be reported in the email log. We also encountered inconsistencies in how data values were displayed. For instance, ``Health \& Wellness'' appeared as ``Health \& Wellness '' with an extra space at the end when the categories were first added to the Zoom Marketplace. This issue has since been fixed, and we updated our dataset to ensure consistency with the new data.

These measures ensured data completeness and accuracy, with particular attention to issues such as failed page loads, missing or incomplete data, and data consistency across different phases of collection. This methodology allowed us to create a comprehensive dataset while maintaining high data quality standards and respecting the technical constraints of the platform.

\subsection{Privacy policy analysis method}


\paragraph{\textbf{Privacy policy collection}.} Our privacy policy analysis methodology is built upon the marketplace data collection infrastructure. Using the Puppeteer library, we had already implemented for marketplace crawling; we extended our automated collection system to handle privacy policy documents. The system was configured to access the privacy policy URLs identified during the initial marketplace crawling phase, maintaining the same 10-second delay between requests to respect server limitations and implementing similar error-handling mechanisms as our main crawler. For each application in our dataset, we visited the previously stored URLs for the corresponding policy page and downloaded it (if the link was valid). For retrials and manual reviews, the framework kept logs of failure cases, e.g., due to non-existent links or any errors due to parsing or network connectivity. For example, if there were a failure in obtaining a privacy policy, the framework would automatically attempt to rerun the HTML download. If the immediate rerun also failed, the framework would add the app to the queue for another attempt in the second pass at the end of the data collection for all apps in the first pass. Apps that still had errors after the second pass were logged for manual verification and reported to researchers via email. We saved the raw HTML content for further analysis and maintained detailed logs of any failed attempts for manual verification and retry procedures.

\paragraph{Automated privacy policy analysis.} To process the collected policy documents, we utilized PoliGraph~\cite{cui2023poligraph}, a specialized natural language processing tool that analyzes unstructured privacy policy texts to create knowledge graphs. PoliGraph identifies statements about data collection and sharing in privacy policies, and builds relationships among data, actors, and actions, such as what data is being collected, who is collecting it, and for what purposes. It then creates knowledge graphs containing nodes and links to represent these relationships. 

\paragraph{Post processing knowledge graphs.} While knowledge graphs are visually rich and ease the process of reviewing and grasping data flows, our ultimate goal was to summarize data collection and sharing statistics. Thus, we developed a Python script to post-process the graph specifications generated by PoliGraph. The script enumerated graph specifications and parsed different relationships (such as generic data types like `contact information' and specific data types such as `phone number'), identified unique data collector entities and purposes and aggregated all these results to compute high-level statistics.  


\paragraph{Ethical concerns.}
There has been a growing recognition among the research community about the need to consider how (measurement) research may adversely impact service providers and how to
minimize those impacts~\cite{redline}. For this study, we exercised strategies to minimize the impact on the servers from which we gathered data, ensuring that normal operations remain unaffected. For example, we implemented appropriate rate limiting to prevent server overload, spacing each call to the Zoom Marketplace by 10 seconds. Additionally, we conducted our weekly data collection between 12 AM and 4 AM on Sundays, when server usage is presumably minimal. We also note that our research can potentially benefit Zoom by helping them identify malicious apps, which can outweigh the computation cost we incurred; we are already in the process of reaching them with our findings. 

\section{Findings}


We present findings to portray the most recent (as of December 2024) status of the Zoom marketplace from privacy and security perspectives, as well as how things (e.g., data access) have changed over the year (from December 2023 to December 2024). To show changes, we compare findings across different time intervals: monthly and half-yearly. For the later, we compare among three time points---December 2023, May 2024, and December 2024. Data about app categories are not available for the first six months since categories were introduced in March 2024; thus, findings that rely on category data are compared between the last two dates. Additionally, we supplement quantitative data with manual reviews of apps and their privacy policies to provide a deeper and nuanced understanding of privacy and security issues.


\subsection{App trend analysis}

\subsubsection{Number of apps over time.} 

\begin{figure}[!h]
    \centering
    \includegraphics[width=0.95\linewidth]{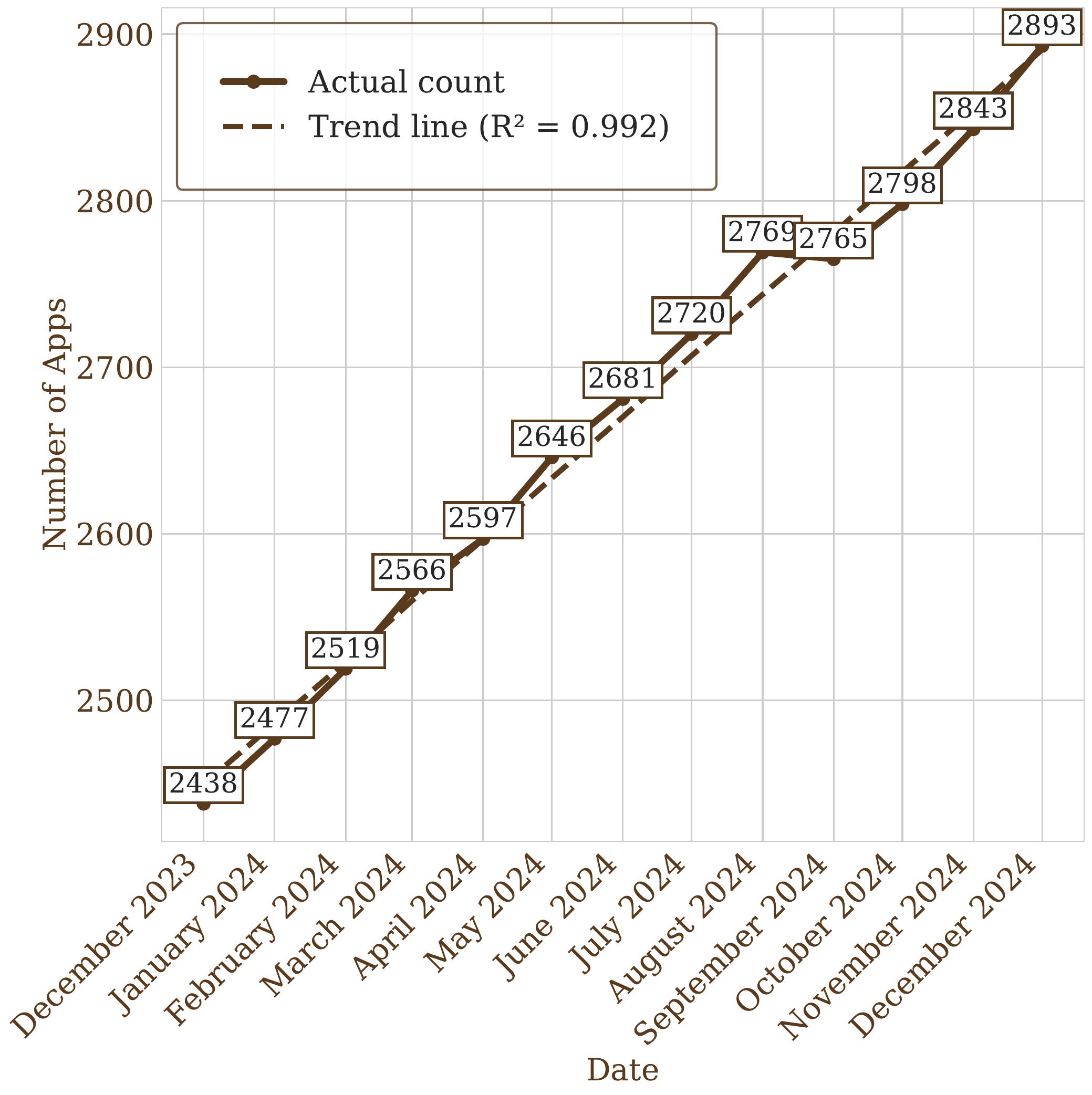}
    \caption{Monthly trend in the total number of apps.}
    \label{fig:app-trend}
\end{figure}

From December 2023 to December 2024, the total number of unique apps changed from 2,438 to 2,893 with a monthly trend of linearly increasing (Figure~\ref{fig:app-trend}). While the trend is upward, a small number of apps were also removed from the marketplace each month. In total, between December 2023 and December 2024, 260 apps were removed, with 162 apps being removed in the second half. The largest number of apps were removed from the Scheduling category (n=21).


\begin{figure*}[!h]
    \centering
    \includegraphics[width=.9\textwidth]{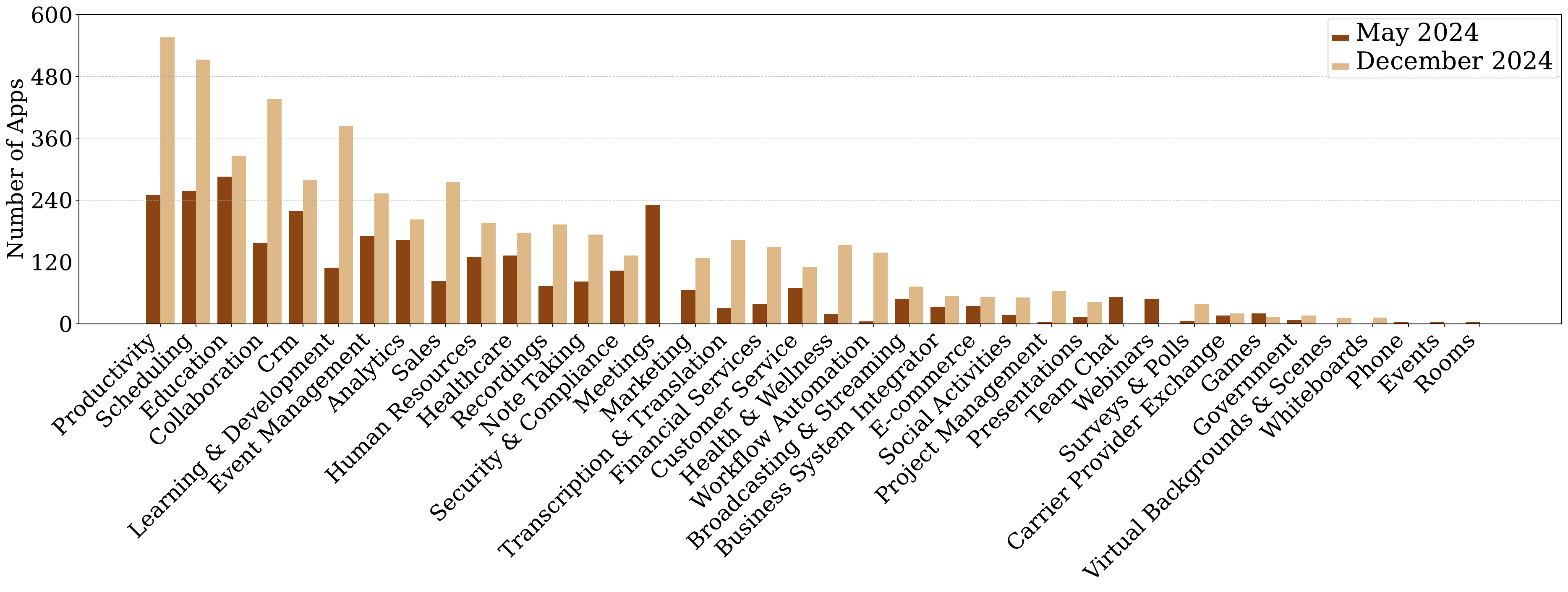}
    \caption{Change in the number of apps per category from May 2024 to December 2024.}
    \label{fig:per-cat-trend}
\end{figure*}

Categories were introduced in March 2024, and as of December 2024, there were 32 categories. The largest number of apps were under \textit{Productivity} (n=556) while \textit{Virtual Backgrounds \& Scenes} had the fewest apps (n=11). The general upward trend in the number of apps was also observed across the categories. As Figure~\ref{fig:per-cat-trend} shows, almost all categories grew in the number of apps between May and December 2024, with a few (such as \textit{Productivity} and \textit{Scheduling}) experiencing relatively much larger growth. 

\subsubsection{Overlaps in app categories and correlation with app features}
An app can be listed under multiple categories, reaching a larger potential user base. Still, this marketplace feature can also be abused to spam users as well as to ask for unnecessary permissions~\cite{seneviratneEarlyDetectionSpam-2015}. Investigating the trend in cross-category overlaps, we found that the number of categories per app dramatically changed between May and December 2024. For example, in May 2024, almost 89.68\% (n=2373) of apps were listed under one category; only 205 and 68 apps were listed under two and three categories, respectively. The \textit{Education} category shared the highest number of apps with other categories: 197 apps with \textit{Learning \& Development}, n=51 apps with \textit{Scheduling}, and 34 apps with \textit{Collaboration}.

\begin{figure*}[!h]
    \centering
    \includegraphics[width=.95\textwidth]{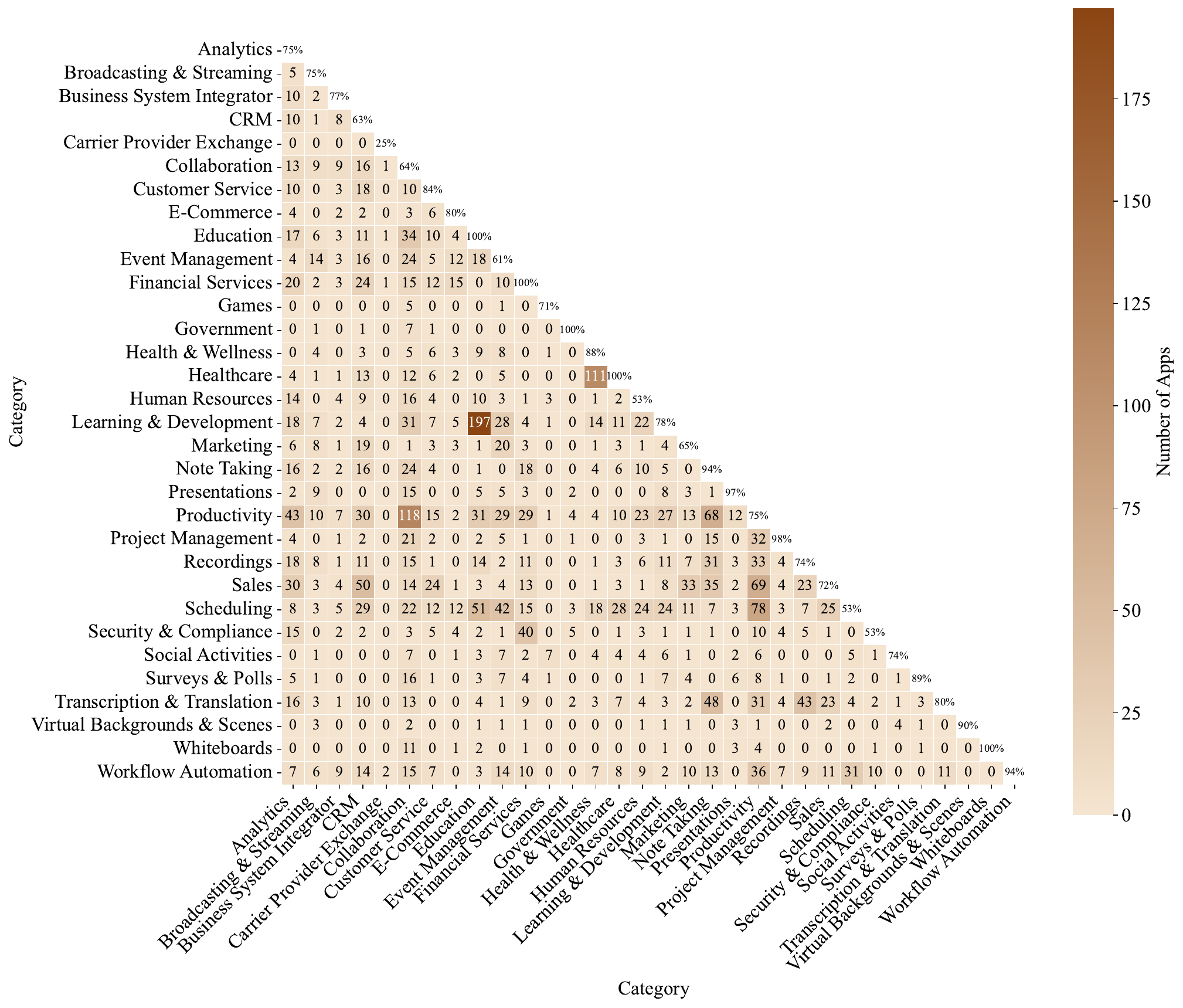}
    \caption{Overlaps in app categories (diagonal cells show the percentage of apps in a category shared with other categories.)}
    \label{fig:cat-heat}
\end{figure*}

In contrast, by December 2024, only 46\% (1344) of the apps were listed in one category; 735 apps had two,  686 apps had three, and 128 had four categories. For example, apps such as Akute (\textit{Health}) and Intellecta (\textit{Education}) were listed under one category in May, but that changed to four categories by December 2024. Figure~\ref{fig:cat-heat} visualizes cross-category overlaps. Generally, there are large overlaps between thematically similar categories, such as \textit{Health} and \textit{Health \& Wellness}, \textit{Education} and \textit{Learning \& Development}, and \textit{Transcription \& Translation} and \textit{Note taking}. However, there are overlaps between seemingly unrelated categories, such as \textit{Customer service} and \textit{Learning \& Development}. By manually reviewing descriptions of 10\% apps from each category, we identified apps that were potentially mis-categorized. For example, WRKiiT Beta provides event management services but was cross-listed under \textit{Health \& Wellness} and \textit{Learning \& Development}. We also found apps that included categories seemingly unrelated to their functionality: YouStudio and Kindred Minds provide remote class and AI-based leadership coaching services, respectively, but both were also listed under \textit{Health \& Wellness}. 

Motivated by the above examples, we next investigated whether the inclusion of new categories in existing apps was accompanied by additional functionality relevant to those new categories. Since app description pages detail app functionality, we examined whether apps included new categories between May and December 2024 and whether their descriptions changed within that time interval. We found that among the 2484 apps that were present in both May and December of 2024, 1356 (55\%) added at least one new category, but only 184 revised their description, and among them, 125 apps changed their description by less than 10\% in terms of unique word count. Manual review of the old and new descriptions of these 125 apps revealed that most of the changes were minor, such as adding or removing white spaces between punctuation and words (which would be treated as a different word now). Several apps appeared to have included categories without providing associated functionality; for example, \textit{Music Player - YouTube, Spotify \& More} from BlueSky Apps streams music, but was listed under \textit{Healthcare} and \textit{Event management} (in addition to \textit{Broadcasting \& Streaming}). We also manually reviewed 20 randomly selected apps that did not update descriptions at all after including new categories and identified apps that were possibly misclassified. For example, \textit{Thalamus}, an interview management program for Graduate Medical Education, was listed under \textit{Healthcare} and \textit{Health \& Wellness} but did not appear to provide any health-related services. These results hint at potential spamming activities, where an app bundles unrelated categories and keywords to appear more frequently in search results and reach more potential users~\cite{seneviratneSpamMobileApps-2017}.

Adding more categories was also accompanied by apparently unnecessary data permissions. For example, Wavoto provides Sales website templates and hosting services but was listed under \textit{Learning \& Development} and \textit{Scheduling}, and requires view and manage access to meeting content and participants' profiles. Moreover, we found that existing apps most commonly added \textit{Learning \& Development} (n=237), followed by \textit{Health \& Wellness} (n=126). Since user data generated in education and health contexts are deemed more sensitive and are protected under FERPA~\cite{ferpa} and HIPAA~\cite{hipaa-definition}, the proliferation of unnecessary categorization and data permission raises privacy compliance concerns.


\subsection{App permission analysis}
\subsubsection{Overall trend.}
Zoom has different permission categories that provide either view (read-only) or manage (edit) access to user data. This data can be associated with only the user who added an app to their Zoom client (User only) or with other people (User and others), such as meeting participants (see Table~\ref{tab:zoom-permissions}).
\begin{table}[]
    \centering
    \renewcommand{\arraystretch}{1.2} 
    \begin{tabular}{>{\raggedright\arraybackslash}p{3cm} >{\raggedright\arraybackslash}p{5cm}} 
    \toprule
    \textbf{Permission Category} & \textbf{Data Accessed} \\
    \midrule
    Profile \& Contact Information (User only) & user name, display name, picture, email address, phone number, job information, stated locale, account, user ID, contact lists  \\
    Product Usage (User and others) & when participants join/leave, whether participants sent messages and who they message, performance data \\
    Settings (User only) & whether a passcode or a waiting room is required, permitted event capacity, screen sharing settings \\
    Content (User and others) & audio, video, messages, transcriptions, feedback, responses to polls and Q\&A, files, invitation details, meeting or chat name, and meeting agenda  \\
    Calendars (User only) & calendar of scheduled Zoom meetings and webinars  \\
    Registration Information & name and contact information, responses to registration questions  \\
    Participant Profile \& Contact Information (User and others)  & name, display name, email address, phone number, user ID  \\
    Functional (User and others) & Zoom user ID, session IDs, meeting role, and information about your meeting, webinar, or chat  \\
    Device Information (User only) & speakers, microphone, and camera, OS version, hard disk ID, PC name, MAC address, IP address and general location at the country level derived from it \\
    Account Information (User only) & administrator name, account email address, billing information, and account plan information  \\
    \bottomrule
    \end{tabular}
    \caption{Permission categories and associated user data. (User only) implies data about the primary user, as opposed to also about other meeting participants.}
    \label{tab:zoom-permissions}
\end{table}

\begin{figure}[!h]
    \centering
    \includegraphics[width=0.95\linewidth]{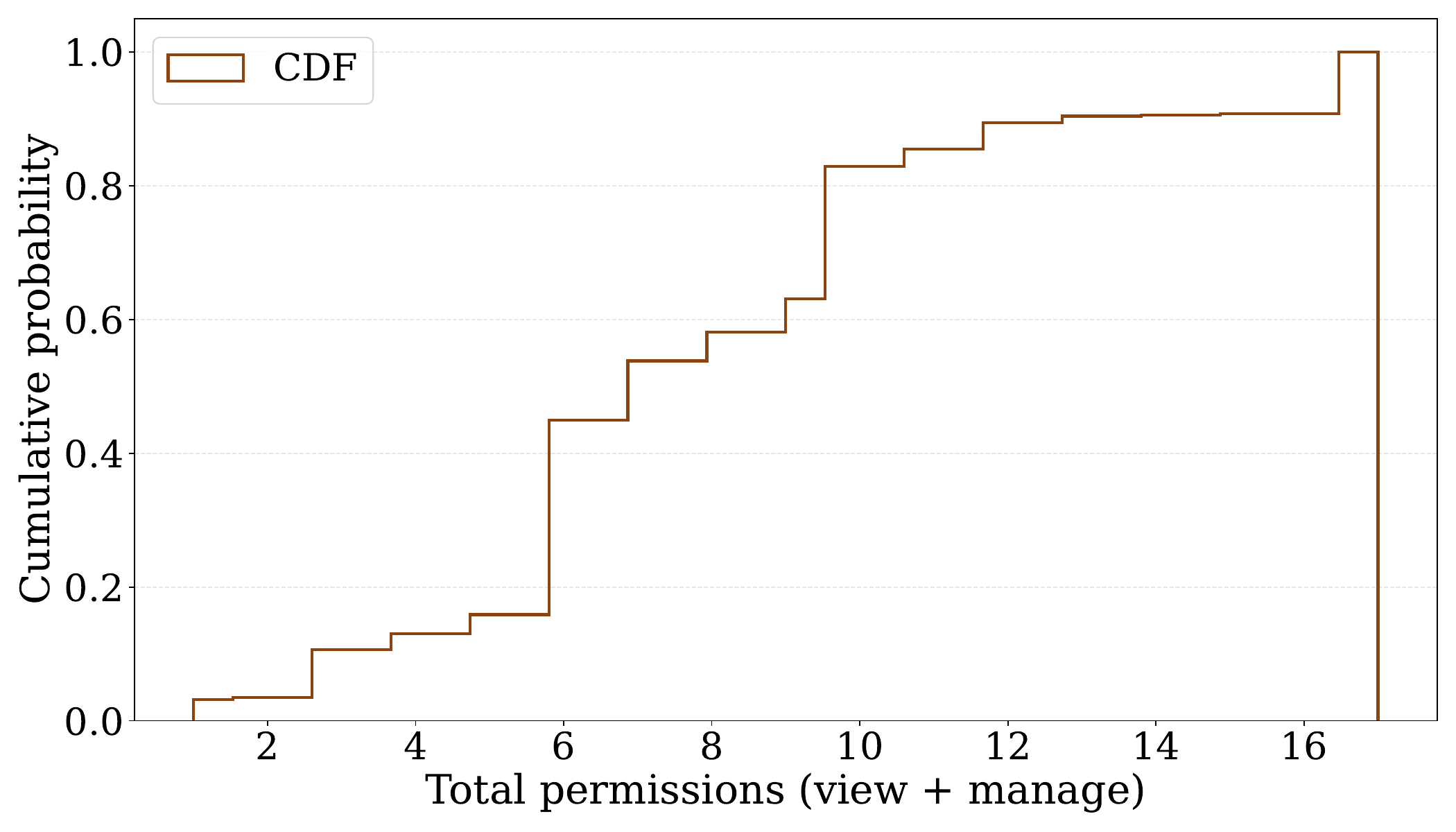}
    \caption{CDF Plot of Total Permissions per App.}
    \label{fig:cdf-all-permissions}
\end{figure}

Figure~\ref{fig:cdf-all-permissions} shows the cumulative distribution function of the total number of permissions (view and manage, combined) per app based on the latest data (collected in December 2024); the majority of the apps require between 6 to 10 permissions. Table~\ref{tab:view-perm} and Table~\ref{tab:manage-perm} list the number of apps that require different view and manage permissions, respectively. For view permissions, almost all apps access the profiles and contact information of the primary users, which in reality may include personal information about other people who are in the users' contact list (Table~\ref{tab:zoom-permissions}). This was closely followed by arguably less sensitive data about product usage and settings. 
In contrast, apps most frequently require manage permissions to meeting content, participants, and registration and scheduling information (Table~\ref{tab:manage-perm}). 

\begin{table}[]
    \centering
    \begin{tabular}{l r r}
    \toprule
    \textbf{View Permission December 2024} & \textbf{Count} & \textbf{Percentage} \\
    \midrule
    Profile \& Contact Information & 2661 & 92.0\% \\
    Product Usage & 2584 & 89.3\% \\
    Settings & 2540 & 87.8\% \\
    Content & 1757 & 60.7\% \\
    Calendars & 1587 & 54.9\% \\
    Registration Information & 1528 & 52.8\% \\
    Participant Profile \& Contact Information & 1494 & 51.6\% \\
    Functional & 452 & 15.6\% \\
    Device Information & 416 & 14.4\% \\
    Account Information & 369 & 12.8\% \\
    \bottomrule
    \end{tabular}
    \caption{View Permission Counts and Percentages}
    \label{tab:view-perm}
\end{table}

\begin{table}[]
    \centering
    \begin{tabular}{l r r}
    \toprule
    \textbf{Manage Permission December 2024} & \textbf{Count} & \textbf{Percentage} \\
    \midrule
    Content & 2080 & 71.9\% \\
    Participants & 1998 & 69.1\% \\
    Registration \& Scheduling & 1938 & 67.0\% \\
    Settings & 568 & 19.6\% \\
    Profile \& Contact Information & 487 & 16.8\% \\
    Account Information & 278 & 9.6\% \\
    Devices & 228 & 7.9\% \\
    \bottomrule
    \end{tabular}
    \caption{Manage Permission Counts and Percentages.}
    \label{tab:manage-perm}
\end{table}

\subsubsection{Permissions analysis across app categories.}
\begin{figure*}[!h]
    \centering
    \includegraphics[width=0.95\linewidth]{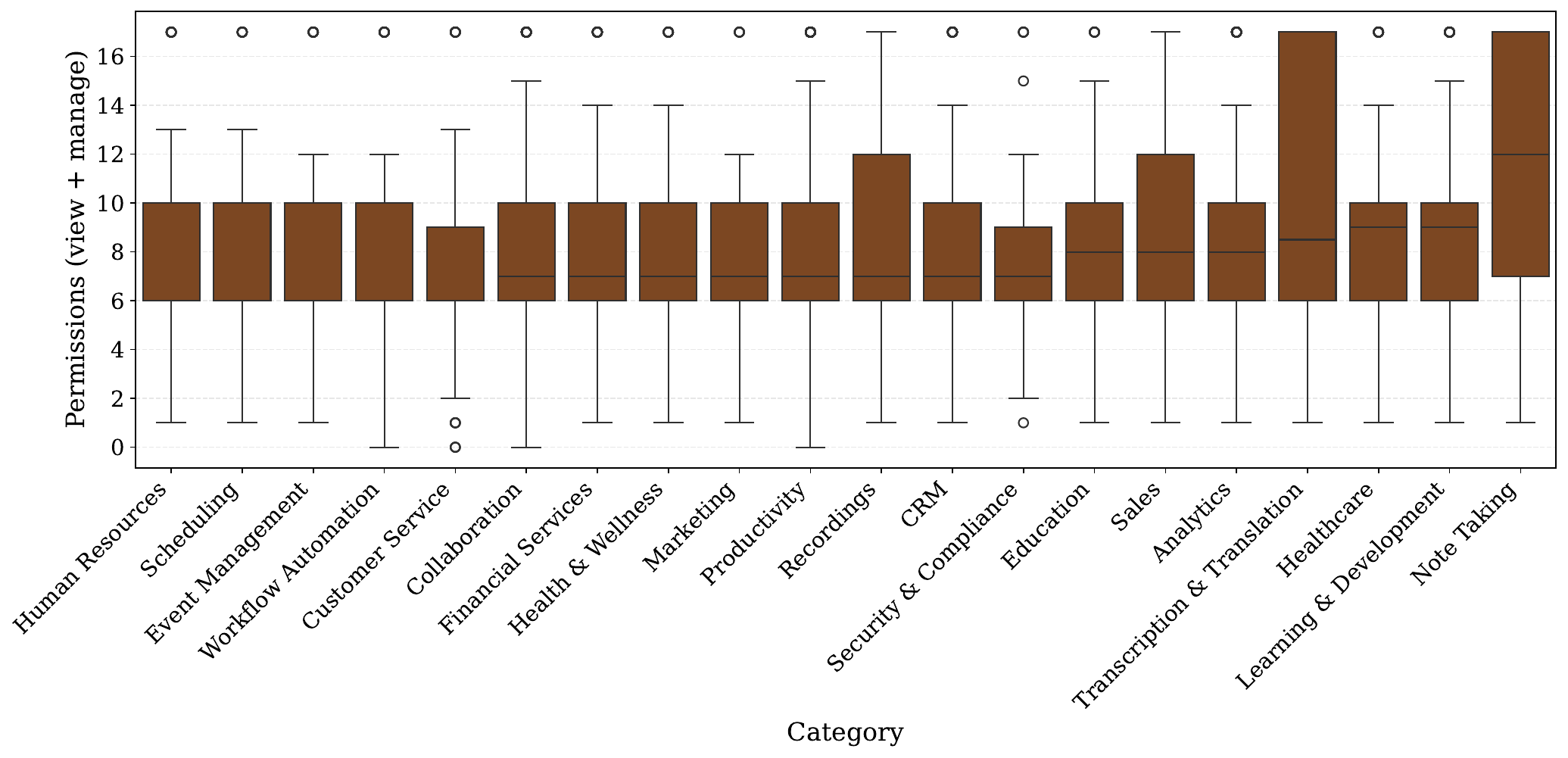}
    \caption{Permission count for top 20 categories (sorted by the median number of permissions).}
    \label{fig:top-20-categories-boxplot}
\end{figure*}

Figure~\ref{fig:top-20-categories-boxplot} shows box plots for the number of permissions required by apps in 20 categories with the largest number of apps (total 2,813 apps, that is 97.23\% of all apps). Interestingly, the cross-category distributions of inter-quartile range look similar, indicating that apps, regardless of what types of functionalities they provide, ask for roughly the same number of permissions (between 6 and 10) except for two categories (\textit{Note Taking} and \textit{Transcriptions and Translations}) that require a slightly higher number of permissions. The overlap in categories (and hence functionality and data requirements) may partly explain this uniformity; however, our manual review of the permissions from two apps in each category revealed that apps indeed ask for data that seem to be unrelated to their functionalities. For example, \textit{Calendly for Zoom}, which automatically creates video conference details and saves them to Calendly event, requests access to meeting content, such as audio, video, and messages, generated by \textit{all} participants. 

\begin{figure*}[!h]
    \centering
    \includegraphics[width=.95\textwidth]{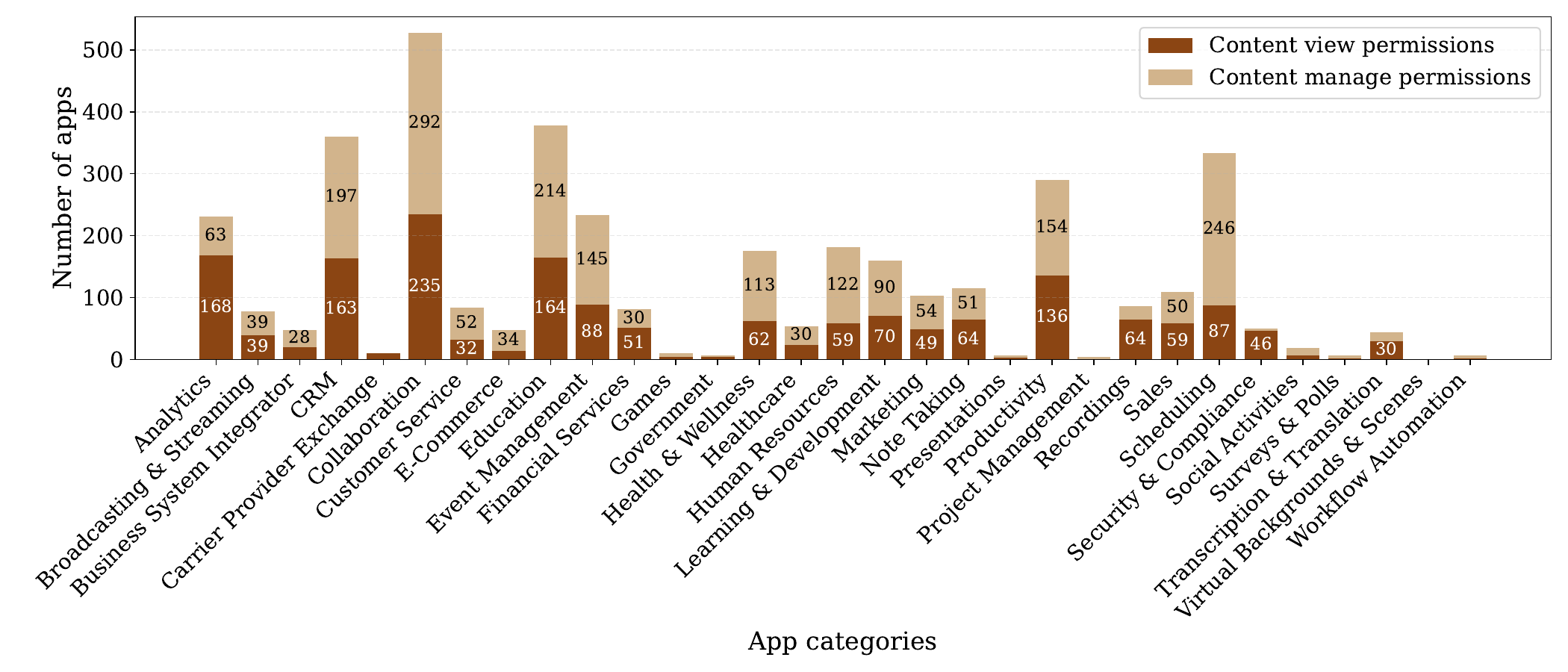}
    \caption{Number of apps in different categories that request to view or manage meeting content.}
    \label{fig:content-permission}
\end{figure*}

\paragraph{Meeting content permissions.}
Content data are arguably the most sensitive as they contain personally identifiable biometric information including voice, facial features, retina, and behavioral patterns; privacy and safety risks from these data have dramatically risen with the ubiquity of technology that can create fake images, audio, or video using them. Thus, we investigate the use of content data more closely by looking at each category, as shown in Figure~\ref{fig:content-permission}. Then, we manually reviewed apps in categories for which the need for meeting content is not readily obvious and identified instances of potentially over-permission requests. For example, scheduling apps like \textit{Calendly}, \textit{Skeding}, and \textit{Leadline Connected Calendar for Zoom} provide services to automate meeting creation and invitation and thus there is no apparent need for them to view or manage content generated during the meeting. Likewise, many E-commerce applications, such as \textit{Calero-SaaS Expense Management} and \textit{Niuco} seem to provide services to manage Zoom licenses, and it's unclear why they require access to meeting content.

\begin{table}[]
    \centering
    \begin{tabular}{p{20mm} r r}
    \toprule
    \textbf{\# Permissions change} & \textbf{\# Apps (View)} & \textbf{\# Apps (Manage)} \\
    \midrule
        -5 & 1 & 0 \\
        -1 & 1 & 0 \\
         1 & 13 & 12 \\
         2 & 5 & 2 \\
         3 & 26 & 20 \\
         4 & 7 & 2 \\
         5 & 1 & 0 \\
         6 & 3 & 0 \\
         7 & 11 & 51 \\
         8 & 1 & 0 \\
         9 & 28 & 0 \\
    \bottomrule
    \end{tabular}
    \caption{View and manage permissions changes in existing Apps from May 2024 to Dec 2024}
    \label{tab:perm-change}
\end{table}

\subsubsection{Changes in permissions requirements over time}\label{sec:perm-change}
Table~\ref{tab:perm-change} shows the number of apps that increased or decreased permissions between May and December 2024. As the table shows, permission requests by existing apps remained relatively stable over time; few apps changed view or manage permissions between May and December 2024. In particular, only two apps (Theta Lake eComms Archive and Biznest-AI Discovery Sidekick) removed permissions, and the total number of apps that added one or more permissions is below 100 for both view and manage permissions.

Apps created after May 2024, however, tended to ask for more permissions than older apps. For example, only 4\% (n=25) of apps in May 2024 asked for all view permissions, whereas 30.56\% (125 out of 409) of apps added between May and December 2024 asked for all view permissions. Likewise, except for 24 apps, no other app that existed in May 2024 required any manage permissions, but 30.56\% (the same 125 apps) of apps created afterward asked for all seven manage permissions. This tendency of requiring all permissions was most prevalent among \textit{Note-taking} apps where  40.46\%  (70 apps) asked for all permissions, followed by \textit{Transcription \& Translation} (28.22\% or 46 apps)---these two categories also had the highest median number of permission requirements (Figure~\ref{fig:top-20-categories-boxplot}). 
A manual review of 10 randomly selected apps from these categories indicates that some apps may be over-privileged. For example, \textit{Gong for Zoom Meetings}, a recording bot, only requires access to meeting content and provides transcription and meeting analysis services, whereas other apps with similar features (e.g., Embra AI Notetaker) require all permissions.

\begin{figure*}
    \centering
    \subfloat[]{
        \includegraphics[width=.45\textwidth]{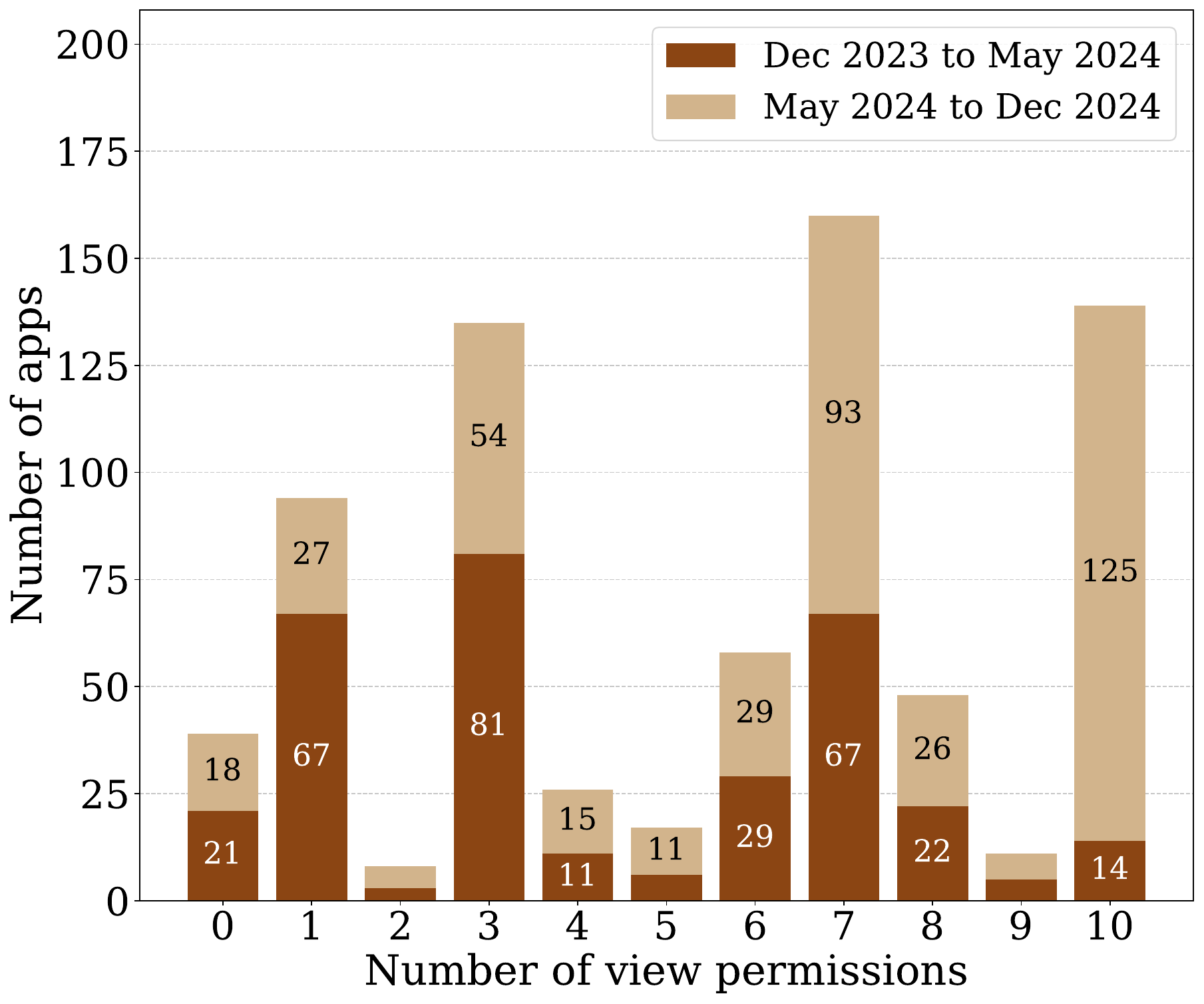}
        \label{fig:view-permission-new-app}
    }
    \hfill
    \subfloat[]{
        \includegraphics[width=.45\textwidth]{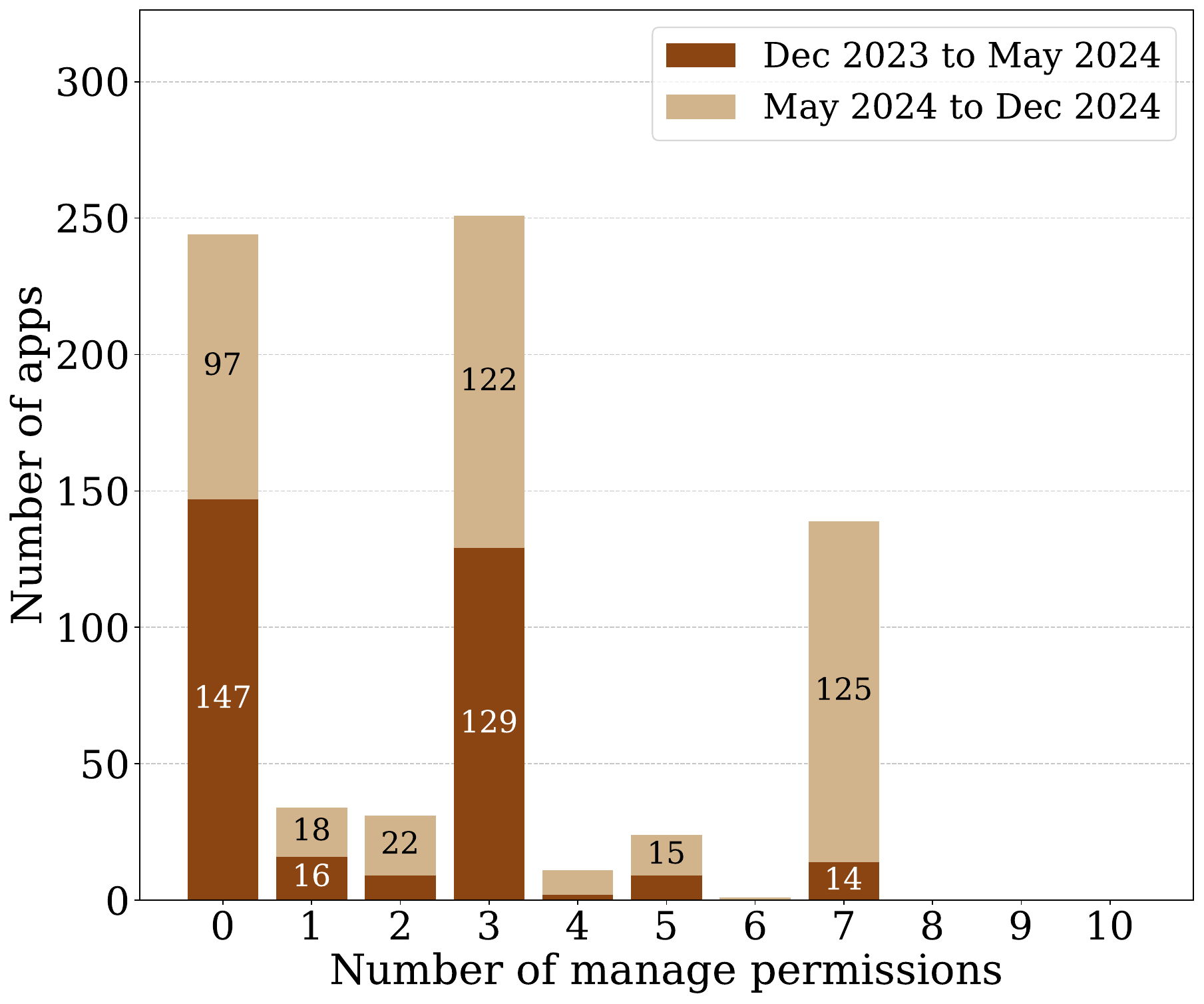}
        \label{fig:manage-permission-new-app}
    }
    \caption{Distribution of view and manage permissions requested by Zoom apps created before and after May 2024.}
    \label{fig:new-app-trend}
\end{figure*}

\subsection{Privacy policy analysis}
We identified and analyzed 1,831 valid app privacy policies in the latest dataset (after December 2024). We could not analyze policies for the remaining 1,079 apps either because of invalid or non-existent links to privacy policies (n=978) or non-English privacy policies.






\subsubsection{Data collection and sharing practices} 
\begin{figure}
    \centering
    \includegraphics[width=0.95\linewidth, angle=0]{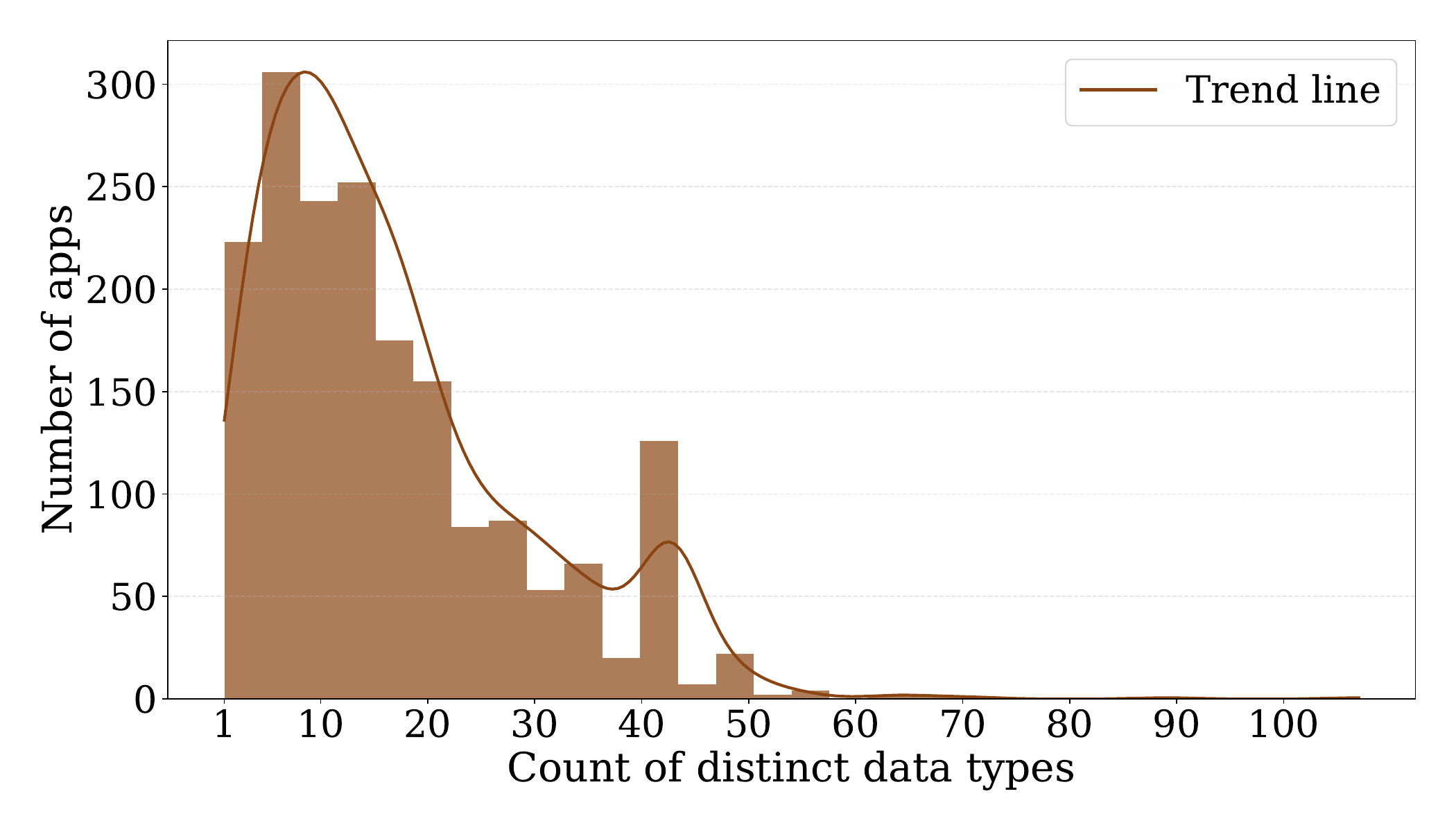}
    \caption{Histogram of the number of distinct data items mentioned in privacy policies.}
    \label{fig:policy-data-count}
\end{figure}

In total, the collected privacy policies contain 43,467 statements about data collection for 7,238 unique data items (e.g., \textit{Phone number} and \textit{Geolocation}). Figure~\ref{fig:policy-data-count} provides high-level trends: e.g., the distribution of the number of data items per policy is highly skewed, where a majority (n=1,230) mention fewer than 20 data items (including 230 only mentioning one data item), a significantly large number of policies (n=118) specify 20--60 data items, and finally a small number (n=38) of them specify more than 100 data items.

Table~\ref{tab:top-data-types-expanded} lists the 20 most frequently mentioned data items. Worryingly, at the top is \textit{UNSPECIFIED\_DATA}, meaning that the statements might have been too vague for Poligraph to determine the specific data that were being referred to. We manually reviewed 20 examples of such cases and found that they included statements revealing data collection practices without specifying what data (e.g., ``We may allow third-party advertising partners to set tracking tools to collect information regarding your activities'' (PocketSuite) 
and `` we may use Google Analytics and other analytics tools such as Fabric.io to collect and process data'' (Ovatu)).

At the second position are Cookie and Pixel tags, also raising privacy concerns as they frequently are used for advertising or other unspecified purposes. Table~\ref{tab:top-data-types-expanded} also contains broad and potentially vague categories, such as \textit{Information about you}, \textit{Internet activity}, and \textit{Non-personal information}; they are accompanied by statements like  ``We use the information we collect or receive [$\cdots$]'' (Staircare) and ``[$cdots$] In general we use the information we collect to provide and administer the Services [$cdots$]''(SalesHood).

\begin{table*}[]
    \centering
    \renewcommand{\arraystretch}{1.2} 
    \begin{tabular}{l r@{\hskip 10mm} r r r r r r}
    \toprule
    \textbf{Data Type} & \textbf{Total} & \multicolumn{6}{c}{\textbf{Purpose}} \\ 
    & & {Services} & {Analytics} & {Advertising} & {Security} & {Legal} & {Unspecified}\\
    \midrule
    UNSPECIFIED\_DATA & 4535 & 1361 & 724 & 657 & 603 & 560 & 2678 \\
    Cookie / Pixel Tag & 2078 & 1246 & 1134 & 744 & 461 & 404 & 497 \\
    Personal Information & 1972 & 959 & 697 & 611 & 396 & 382 & 868 \\
    Email Address & 1293 & 902 & 722 & 608 & 516 & 471 & 321 \\
    Person Name & 969 & 677 & 535 & 448 & 400 & 341 & 240 \\
    IP Address & 835 & 505 & 472 & 353 & 283 & 247 & 247 \\
    Geolocation & 677 & 517 & 392 & 307 & 407 & 296 & 91 \\
    Contact Information & 600 & 499 & 336 & 290 & 312 & 241 & 63 \\
    Information About You & 598 & 369 & 296 & 242 & 208 & 190 & 192 \\
    Aggregate / Deidentified / \\ Pseudonymized Information & 515 & 203 & 177 & 114 & 80 & 77 & 243 \\
    File & 392 & 114 & 106 & 17 & 15 & 9 & 273 \\
    Phone Number & 370 & 327 & 272 & 248 & 212 & 195 & 21 \\
    Internet Activity & 357 & 205 & 180 & 169 & 129 & 97 & 104 \\
    Postal Address & 327 & 280 & 236 & 205 & 188 & 185 & 34 \\
    Credit / Debit Card Number & 289 & 143 & 98 & 77 & 94 & 61 & 137 \\
    Personal Identifier & 264 & 178 & 149 & 110 & 93 & 48 & 71 \\
    Information We Collect & 253 & 192 & 155 & 138 & 121 & 114 & 54 \\
    Non-Personal Information & 249 & 144 & 136 & 116 & 86 & 91 & 52 \\
    Usage Information & 219 & 214 & 124 & 121 & 204 & 119 & 3 \\
    Browser Type & 201 & 125 & 101 & 64 & 65 & 65 & 50 \\
    Identifier & 198 & 118 & 89 & 81 & 116 & 113 & 30 \\
    Voiceprint & 183 & 180 & 177 & 91 & 91 & 90 & 3 \\
    \bottomrule
    \end{tabular}
 \caption{Top 20 data items and the purposes for their collection. Note that each item can have multiple purposes.}
    \label{tab:top-data-types-expanded}
\end{table*}

Privacy policies may refer to data at a different level than what users see in permission dialog (contact information can refer to either phone number, physical or email address, or all of these). 
To understand if all the data items mentioned in privacy policies are visible to users (through permission prompts), we manually reviewed the text corresponding to the top 100 data items mentioned in privacy policies and mapped them to permissions that are visible to users (Table~\ref{tab:policy-permission-mapping} in \S~\ref{sec:data-mapping}). We also manually reviewed 10 randomly selected privacy policies. We found that most data items can be mapped to permissions, but there are exceptions since developers can collect data through other means. For example, the privacy policies of several apps (e.g., \textit{read.ai} \textit{Insight LMS}, \textit{5mins.ai}) state that they may use the data collected to derive new information related to demographics, employment, and behavioral characteristics. Moreover, some developers state that they might sell the collected or derived data to brokers and other third parties (e.g., Warmly, the developer of \textit{Nametags}). We also found that developers may collect data about their users from other sources, including business partners, social media, other customers, and data brokers (e.g., people.ai and read.ai), and these data items do not correspond to permission prompts and may go unnoticed by the users.


\paragraph{Data collection purposes.}

About 29.32\% (n=12,744) of data statements about data collection did not have any specific purpose stated in the privacy policy. Table~\ref{tab:data-purpose} shows their purposes for the remaining statements. Looking at the app level, almost all (95.68\%) apps mentioned at least one purpose behind collecting data; the remaining apps did not specify any purpose for any of the data items they collected. 

\begin{table}[]
    \centering
    \begin{tabular}{l r}
    \toprule
    \textbf{Purpose} & \textbf{Count} \\
    \midrule
    Services & 4494 \\
    Analytics & 3371 \\
    Advertising & 2719 \\
    Security & 2210 \\
    Legal & 1810 \\
    \bottomrule
    \end{tabular}
    \caption{Purposes for Data Collection}
    \label{tab:data-purpose}
\end{table}

Table~\ref{tab:data-collectors} shows the most frequently mentioned entities that collect or receive user data; unspecified actors are the most common recipients after the first party (i.e., developers). The privacy policies of 93.83\% of apps mentioned that they share at least one data item with third parties. There were 18,862 such data-sharing statements, but only 42.26\% (n=7972) were accompanied by any purpose.  

 \begin{table}[]
    \centering
    \begin{tabular}{l r}
    \toprule
    \textbf{Collector} & \textbf{Count} \\
    \midrule
    We (The app or its developer) & 1787 \\
    UNSPECIFIED\_ACTOR & 1331 \\
    Service Provider & 360 \\
    Advertiser & 304 \\
    Google & 237 \\
    Analytic Provider & 187 \\
    Social Media & 150 \\
    Business Partner & 138 \\
    Zoom & 94 \\
    Invitee & 93 \\
    Integration & 90 \\
    Meeting Host & 88 \\
    \bottomrule
    \end{tabular}
    \caption{Data collectors and recipients}
    \label{tab:data-collectors}
\end{table}


\subsubsection{Change in privacy policies over time} 
Similar to permission requests (\S~\ref{sec:perm-change}), we investigated if privacy policies have changed over time and if this change correlates with permission requests. The datasets in the two time points (before May and after December 2024) had 1,119 common privacy policies, 63\% (n=705) of them were updated, while the rest 414 remained unchanged. The number of unique data items in these 1,119 policies increased from 4,985 to 5,217. 
This increase seems commensurate with the slight rise in permission requirements (Table~\ref{tab:perm-change}). 

A closer look revealed that 32\% (n=358) of policies added at least one new data collection statement, while 30\% (n=331) removed at least one such statement. The three most frequently added data items were voice prints (n=261), audio transcripts (n=129), and meeting content (n=129). On the other hand, top removed data items were account information (n=66), feedback provided to Zoom about product ownership (n=64), and website virtual chat data (n=64), which might indicate a reduced commitment to incorporating user input into product development and privacy practices.

Transparency in data collection also did not change much: in May, 70\% of 26,295 data collection statements specified a purpose, while in December, it increased to 71\% (of a total of 28,000 statements). The number of statements on sharing the collected data with third parties that specified the purpose of sharing remained constant at 42\%.

\subsubsection{Data items in privacy policies across app categories}
\begin{figure*} 
    \centering
    \includegraphics[width=0.95\linewidth, angle=0]{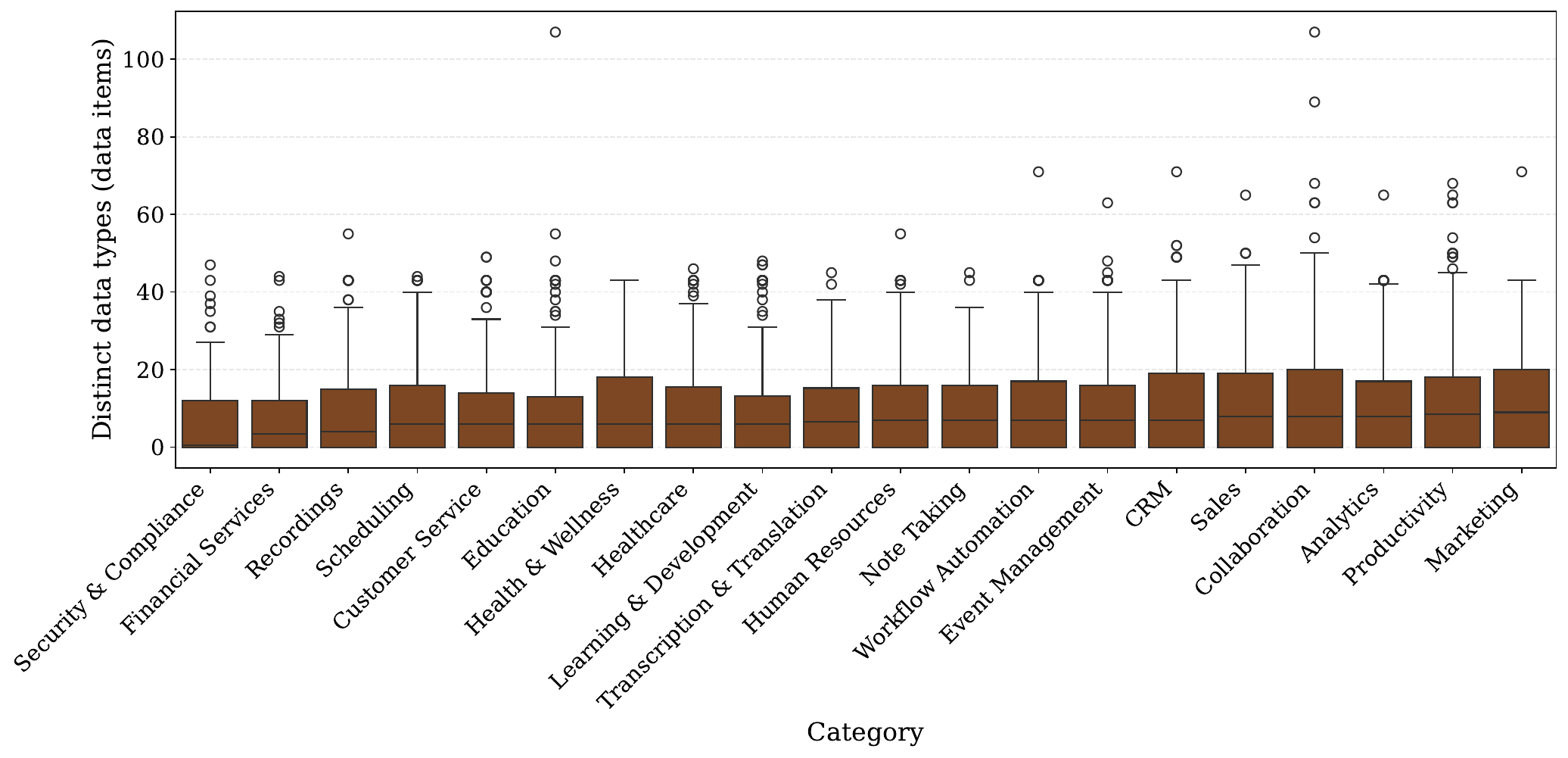}
    \caption{Distribution of the number of unique data items in the privacy policies of apps in the top 20 categories.}
    \label{fig:distinct-data-type-for-top-20-categories-boxplot}
\end{figure*}

Figure~\ref{fig:distinct-data-type-for-top-20-categories-boxplot} shows the number of data items mentioned in the policies for apps in the top 20 categories (in terms of the number of apps). Compared with Figure~\ref{fig:top-20-categories-boxplot}, it seems that while apps in different categories are uniform in data permission requirements, the distributions of data collection disclosure statements vary a lot across categories, with most categories having a lot of outliers. Of particular note, the three categories with the highest number of median permissions: \textit{Note taking}, \textit{Learning \& Development}, and \textit{Healthcare} are likely less transparent in their privacy policies than other categories. This disconnect between data access and disclosure was supported by the result that the number of permissions an app requires was uncorrelated with disclosures and data collection statements in its privacy policy ($r=0.025$, $p>.05$). 

\subsubsection{Health and education apps}
We pay special attention to apps providing health and education services as the data created in those contexts are subject to additional regulatory protections such as FERPA~\cite{ferpa}, HIPAA~\cite{hipaa-definition}, and COPPA~\cite{coppa}. As mentioned in~\S~\ref{sec:background}, apps can either directly comply with these laws or do so through the contracts they make with clients. To determine compliance, we first searched for related keywords (like ``HIPAA'') in the descriptions and privacy policies of apps in the education and health categories. 

Among 142 apps \textit{Healthcare} and \textit{Health and Wellness} categories, 74\%(n=105) did not mention HIPAA. Five of the remaining apps delegated the responsibility for HIPAA compliance to their clients (i.e., healthcare providers and users). For example, IntakeQ, an intake form management service, specifically states that the use and disclosure of protected health information is ``governed by your Provider’s terms and conditions and privacy practices'' while the other 32 apps explicitly mentioned that they are HIPAA compliant (two of them are developed by Zoom).

 Similar analysis of 346 apps in the \textit{Education} and \textit{Learning \& development} categories again revealed three compliance patterns: 88.4\%(n=306) of the apps did not mention either FERPA or COPPA, 5.2\%(n=18) were compliant with both COPPA and FERPA, and another 6.4\%(n=22) of the apps were compliant with only COPPA. Notably, FERPA-compliant apps were more likely to be directly integrated with educational institutions and student information systems as opposed to being used as standalone apps.

   
 
\subsection{Limitations}
Our methodologies have several limitations. First, we relied on automation to scale our analyses, e.g., using PoliGraph to analyze privacy policies. While it is state-of-the-art in this area and outperforms earlier tools by large margins with 97\% precision~\cite{cui2023poligraph}, it still may not detect everything correctly. Our analysis was restricted to the US market and English privacy policies, thus results may not generalize to other countries. We also cannot establish the impact of potential privacy violations, as data about app downloads and user reviews are not public, though it can be safely assumed that the effect is significant since Zoom has been the most popular remote communication tool for the last few years. Finally, while we examined compliance with regulations based on privacy policies, determining who is and is not covered under them requires extensive analysis from legal perspectives and out of scope for this paper, and past research has shown that loopholes can be exploited to bypass regulations or superficially comply with them~\cite{sins-omission, whose-data}. However, we note that privacy and safety risks from improper data collection practices remain the same regardless of whether the collector is covered under privacy regulations. 
\section{Discussions and conclusions}

This paper provides a comprehensive picture of how the Zoom marketplace evolved in one year, from the perspective of data privacy and security. As the marketplace has been continuously growing, we also observed concerning practices such as mis- or over-categorization of apps, likely to attract more users, over collection of user data that are apparently unrelated to the provided features, and not being transparent in explaining why some data was collected. While the trend of increasing data collection by apps over time was observed in other marketplaces~\cite{eduMeasuringAlexaSkills2022, weiPermissionEvolutionAndroid-2012}, this was not the case for the Zoom marketplace: we observed that existing apps rarely increased the number of permissions, rather newer apps required more data access. Notably, a large number of data collection related statements in privacy policies include broad or vague classes of data, which may preclude holding the responsible entity accountable in case of privacy invasions.



Perhaps unsurprisingly, most apps share data with third-parties (including selling or renting data, as the privacy policy of read.ai states), including advertisers and other companies known for extensive online tracking and surveillance activities, which is consistent with past research on other platforms~\cite{andowActionsSpeakLouder-2020}. However, the privacy and safety harms from such practices can be more severe in this case, given that the most popular apps serve in education and healthcare domains and the user base includes minors. When services like Zoom are institutionally procured, contracts typically restrict data use and sharing; however, past research has shown that a long chain of vendors and sub-vendors makes it extremely difficult, if not impossible, to keep track of data use or hold entities accountable in case of privacy violations~\cite{edtech-ccs2024}. Use of these apps with personal accounts by educators for teaching purposes has also become ubiquitous~\cite{kelso2025investigating}; as there is no institutional contract restricting data use, and most apps do not directly comply with FERPA or COPPA, such uses raise severe privacy concerns.

This situation is further complicated by inferring new data. For example, \textit{Insight LMS}, a learning management system app, states in its privacy policy that they may use the collected data to infer other information and profile users. Past research has shown that interaction data from such LMS tools can be used to predict demographic attributes, such as gender and age group~\cite{edtech-pets22}. Apps that access audio or video data can infer much more: recent machine learning models can be used to predict many demographics, behavioral patterns and affective status, as well as physical and cognitive disabilities---most of which are protected under FERPA~\cite{ferpa} and HIPAA~\cite{hipaa-definition}. In addition to raising legal compliance issues, such practices also raise ethical concerns, since these data can lead to discrimination and even incrimination of people (e.g., those who are gender non-conforming) in certain states. 

Last but not the least, as Zoom apps collect data not only about their users but also other people whose contact or profile information is within Zoom's reach. This raises interdependent privacy concerns~\cite{idp-survey}, which is prevalent in many other domains~\cite{liuInterdependentPrivacyMobileApp2022}. Capabilities like contact sharing dramatically expand the number of people who can be brought under surveillance by malicious entities.

To conclude, our research surfaced potential privacy and security issues in this emerging marketplace. While many of these issues were identified in other marketplaces, the use cases for Zoom marketplace apps, the prevalence of apps that use "AI," and the relatively easy access to users' visual and voice fingerprints may mean that the consequences of privacy violations may be much more severe than in other contexts. We hope our research will motivate and inform future research on this platform to alleviate the privacy and safety risks to users, and help improve Zoom marketplace policies and app review process.




\appendix
\clearpage
\section{Appendix}
\subsection{Trend in Permission Requests}\label{sec:perm-trend}


\begin{figure*}[h]
    \centering
    \begin{subfigure}{.5\textwidth}
        \centering
        \includegraphics[width=.8\linewidth]{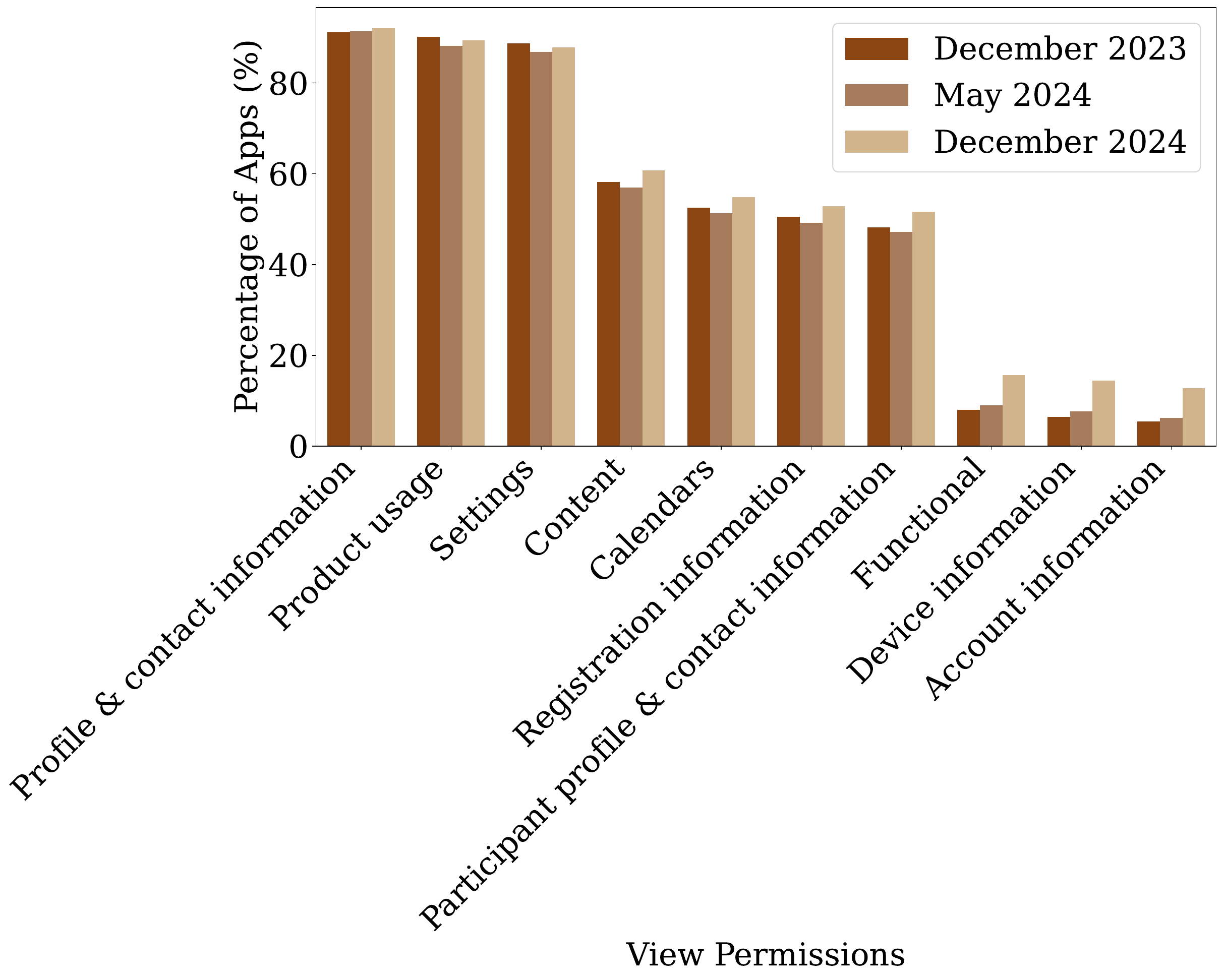}
        \caption{View permissions}
        \label{fig:view-perm-trend}
    \end{subfigure}%
    \begin{subfigure}{.5\textwidth}
        \centering
        \includegraphics[width=.8\linewidth]{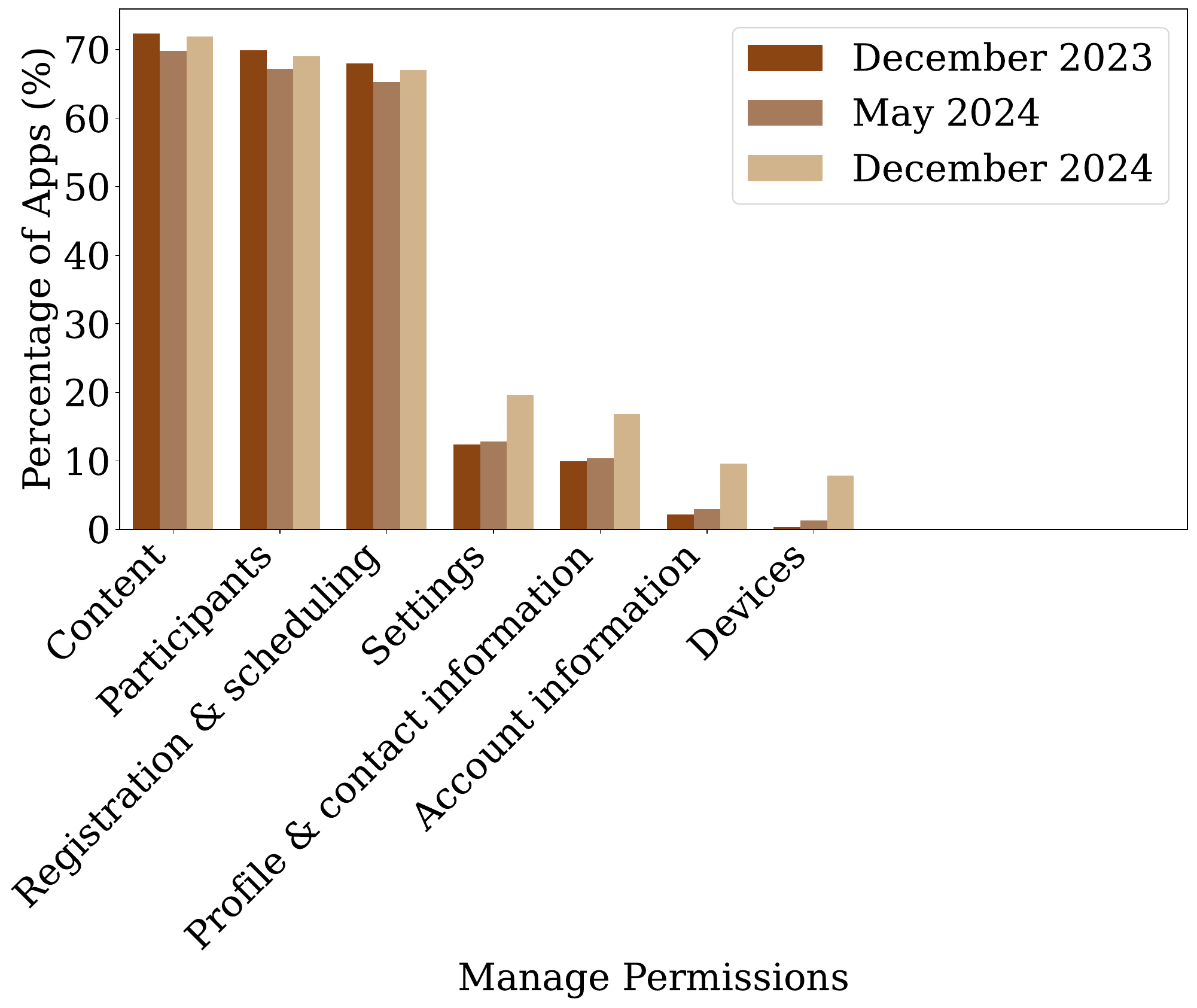}
        \caption{Manage permissions}
        \label{fig:manage-perm-trend}
    \end{subfigure}
    \caption{Percentage of apps with permissions at different time points.}
    \label{fig:permissions-trend}
\end{figure*}


\subsection{Data Items}\label{sec:data-mapping}

\begin{table*}[t] 
    \centering
    \resizebox{\textwidth}{!}{%
    \begin{tabular}{>{\raggedright\arraybackslash}p{3cm} r >{\raggedright\arraybackslash}p{3cm} | >{\raggedright\arraybackslash}p{3cm} r >{\raggedright\arraybackslash}p{3cm}}
    \toprule
        \textbf{Data} & \textbf{Count} &  \textbf{Permission} & \textbf{Data} & \textbf{Count} &  \textbf{Permission} \\ 
    \midrule
    UNSPECIFIED\_DATA & 4535 & - & whiteboard & 173 & Content \\
    cookie / pixel tag & 2078 & Product Usage & meeting content & 173 & Content \\
    personal information & 1972 & Profile and Contact Information & message send to everyone & 173 & Content \\
    email address & 1293 & Profile and Contact Information & configuration information & 172 & Settings \\
    person name & 969 & Profile and Contact Information & audio setting & 172 & Settings \\
    ip address & 835 & Device Information & information on device about face & 172 & Profile and Contact Information \\
    geolocation & 677 & Device Information & information from zoom email services & 172 & Profile and Contact Information \\
    contact information & 600 & Profile and Contact Information & message send to meeting group chat & 172 & Content \\
    information about you & 598 & Profile and Contact Information & message send to they & 172 & Content \\
    aggregate / deidentified / pseudonymized information & 515 & - & information about & 157 & - \\
    file & 392 & Content & device identifier & 154 & Device Information \\
    phone number & 370 & Profile and Contact Information & browsing / search history & 144 & Product Usage \\
    internet activity & 357 & Product Usage & time & 133 & Device Information, Calendar \\
    postal address & 327 & Profile and Contact Information & commercial information & 129 & - \\
    credit / debit card number & 289 & Account Information & password & 126 & Account Information \\
    personal identifier & 264 & Profile and Contact Information & sensitive personal information & 119 & Profile and Contact Information \\
    information we collect & 253 & Product Usage & company name & 117 & Profile and Contact Information, Account Information \\
    non-personal information & 249 & - & title & 115 & Profile and Contact Information \\
    usage information & 219 & Product Usage & metadata & 114 & Product Usage \\
    browser type & 201 & Device Information & registration information & 111 & Registration Information \\
    identifier & 198 & Profile and Contact Information & payment information & 105 & Account Information \\
    voiceprint & 183 & Profile and Contact Information & audio recording & 99 & Content \\
    information about purchase & 176 & Account Information & information browser send & 98 & - \\
    information from partner & 174 & Account Information & calendar & 96 & Calendars \\
    audio transcript & 173 & Content & voice & 95 & Content \\
    \bottomrule
    \end{tabular}%
    }
    \caption{The most frequent 100 data items mentioned in privacy policies, along with the total number of times they are mentioned, and which Zoom permission they map to (`-'' indicates no mapping was possible or uncertain).}
    \label{tab:policy-permission-mapping}
\end{table*}

\clearpage 

\begin{table*}[t] 
    \centering
    \resizebox{\textwidth}{!}{%
    \begin{tabular}{>{\raggedright\arraybackslash}p{3cm} r >{\raggedright\arraybackslash}p{3cm} | >{\raggedright\arraybackslash}p{3cm} r >{\raggedright\arraybackslash}p{3cm}}
    \toprule
        \textbf{Data} & \textbf{Count} &  \textbf{Permission} & \textbf{Data} & \textbf{Count} &  \textbf{Permission} \\ 
    \midrule
    education information & 94 & Profile and Contact Information & datum about you & 73 & Profile and Contact Information \\
    number & 94 & - & information collect & 73 & - \\
    inference & 94 & - & Google & 72 & - \\
    username & 93 & Account Information & device information & 72 & Device Information \\
    date & 92 & Device Information, Calendars & specific information & 70 & - \\
    industry & 88 & Profile and Contact Information & audio information & 70 & Content \\
    Facebook & 87 & - & operating system & 70 & Device Information \\
    information regarding meeting invitation & 87 & Calendars & content & 67 & Content \\
    email metadata used for basic email delivery & 87 & Functional, Content & datum collect & 66 & - \\
    customer record information & 65 & - & payment datum & 65 & Account Information \\
    system log & 53 & Device Information & device type & 52 & Device Information \\
    \bottomrule
    \end{tabular}%
    }
    \caption{Continuation of Table~\ref{tab:policy-permission-mapping}.}
    \label{tab:policy-permission-mapping-cont}
\end{table*}

\end{document}